\documentclass[prl,showpacs,twocolumn]{revtex4}
\usepackage{graphicx}
\usepackage{bm}
\usepackage{pifont}
\usepackage{amsmath}

\begin{document}

\title{Thermal generation of spin current in an antiferromagnet}

\author{S. Seki$^{1,2}$}
\author{T. Ideue$^{3}$} 
\author{M. Kubota$^{1,4}$}
\altaffiliation{Present Address: Corporate Technology and Business Development Unit, Murata Manufacturing Co., Ltd., Nagaokakyo, Kyoto 617-8555, Japan}
\author{Y. Kozuka$^{3}$}
\author{R. Takagi$^{1}$}
\author{M. Nakamura$^{1}$}
\author{Y. Kaneko$^{1}$} 
\author{M. Kawasaki$^{1,3}$}
\author{Y. Tokura$^{1,3}$} 
\affiliation{$^1$ RIKEN Center for Emergent Matter Science (CEMS), Wako 351-0198, Japan}
\affiliation{$^2$ PRESTO, Japan Science and Technology Agency (JST), Tokyo 102-8666, Japan}
\affiliation{$^3$ Department of Applied Physics and Quantum Phase Electronics Center (QPEC), University of Tokyo, Tokyo 113-8656, Japan}
\affiliation{$^4$ Research and Development Headquarters, ROHM Co., Ltd., Kyoto 615-8585, Japan}

\begin{abstract}

Longitudinal spin Seebeck effect has been investigated for an uniaxial antiferromagnetic insulator Cr$_2$O$_3$, characterized by a spin-flop transition under magnetic field along the $c$-axis. We have found that temperature gradient applied normal to Cr$_2$O$_3$/Pt interface induces inverse spin Hall voltage of spin current origin in Pt, whose magnitude turns out to be always proportional to magnetization in Cr$_2$O$_3$. The observed voltage shows significant enhancement for the lower temperature region, which can be ascribed to the phonon-drag effect on the relevant spin excitations. The above results establish that antiferromagnetic spin waves with high frequency above 100 GHz can be an effective carrier of spin current.

\end{abstract}
\pacs{75.30.Ds, 75.70.-i, 65.40.-b}
\maketitle

Spin current, i.e. a flow of spin angular momentum or magnetic moment, has recently attracted revived attention as the potential alternative to charge current with improved energy efficiency\cite{SC_Review, SpinCurrentMurakami, SpinHallExp, SpinHallTheory}. Spin-polarized conduction electrons in metallic systems, as well as spin waves in insulating systems, are considered as the two important carriers of spin current\cite{YIG_SP, YIG_SSE}. Especially, the latter spin-wave spin current (SWSC) has much longer decay length and can avoid the simultaneous flow of charge current accompanied with Joule heat loss, which are strong advantages for the spintronics applications.

In case of ferro/ferrimagnetic insulators (FMI), SWSC can be generated by various external stimuli such as magnetic resonance\cite{YIG_SP, Py_SpinPump} or application of temperature gradient $\nabla T$\cite{YIG_SSE, UchidaNature, UchidaReview, FirstLongSSE, YIG_SSE_NoProx, Theory_SSE}. The latter process is called spin Seebeck effect, and the simultaneous application of $\nabla T$ and magnetic field $H$ to FMI induces SWSC carrying spin angular momentum $\vec{\sigma}$ $(\parallel \vec{H})$. When paramagnetic metal (PM) is attached to FMI, the spin current $\vec{J}_s$ flowing normal to their interface plane is injected into the PM layer through the interfacial spin-exchange interaction\cite{UchidaReview, FirstLongSSE, YIG_SSE_NoProx, Theory_SSE}. This causes inverse spin Hall effect and associated electric voltage $V_{\rm{ISHE}}$ in PM, which is given by 
\begin{equation}
\vec{V}_{\rm ISHE} \propto L_{V} \theta_{\rm SH} (\vec{J}_s \times \vec{\sigma}).
\label{ISHE}
\end{equation}
Here, $\theta_{\rm SH}$ is the spin-Hall angle of PM, and $L_V$ is the gap distance between the electrodes for the voltage measurement. This process can be viewed as a kind of thermoelectric conversion with its efficiency scaling with the film size $L_V$, which may offer an unique route for waste heat utilization without requiring a series connection of thermocouples\cite{NEC_SSE}.

Previously, the studies of SWSC have mainly focused on a limited number of ferrimagnetic insulators\cite{UchidaReview} such as rare-earth iron garnet $R_3$Fe$_5$O$_{12}$ (including Y$_3$Fe$_5$O$_{12}$ (YIG))\cite{YIG_SSE} and spinel ferrite $M$Fe$_2$O$_4$ ($M$: $3d$ transition metal)\cite{Spinel_SSE}. However, the most of magnetic insulators are rather antiferromagnetic\cite{RMP_Tokura}, and it is a crucial issue whether the spin waves in antiferromagnets can carry spin current or not. Since the dynamics of antiferromagnets are characterized by two or three orders of magnitude higher frequency than those of ferromagnets\cite{AFMR_Theory, AFMR_Text}, the antiferromagnetic spin wave can potentially serve as the medium for the ultrafast information processing and communications. Antiferromagnets are also free from stray fields in the ground state, which implies that their dynamics are relatively robust against magnetic perturbations or defects. In general, antiferromagnetic spin wave is described as the propagating precession of two oppositely aligned sub-lattice magnetic moments\cite{AFMR_Theory, AFMR_Text}. Recent theoretical studies have suggested that such local spin oscillations are represented by two degenerated magnon branches carrying opposite sign of spin angular momentum in the limit of $H \rightarrow 0$\cite{AF_SpinPump}, where the total spin current will cancel out for the thermal excitation process\cite{AF_SSE}.

In this Letter, we report the experimental observation of spin Seebeck effect for an uniaxial antiferromagnetic insulator Cr$_2$O$_3$. By applying temperature gradient normal to the Cr$_2$O$_3$/Pt interface, the inverse spin-Hall voltage of spin-current origin has successfully been detected in the Pt layer. The magnitude of thermally induced spin current turns out to be proportional to magnetization in Cr$_2$O$_3$ even under the $H$-induced spin-flop transition, proving that antiferromagnetic spin wave characterized by high frequency above a hundred gigahertz can be an effective carrier of spin current.

\begin{figure}
\begin{center}
\includegraphics*[width=7.5cm]{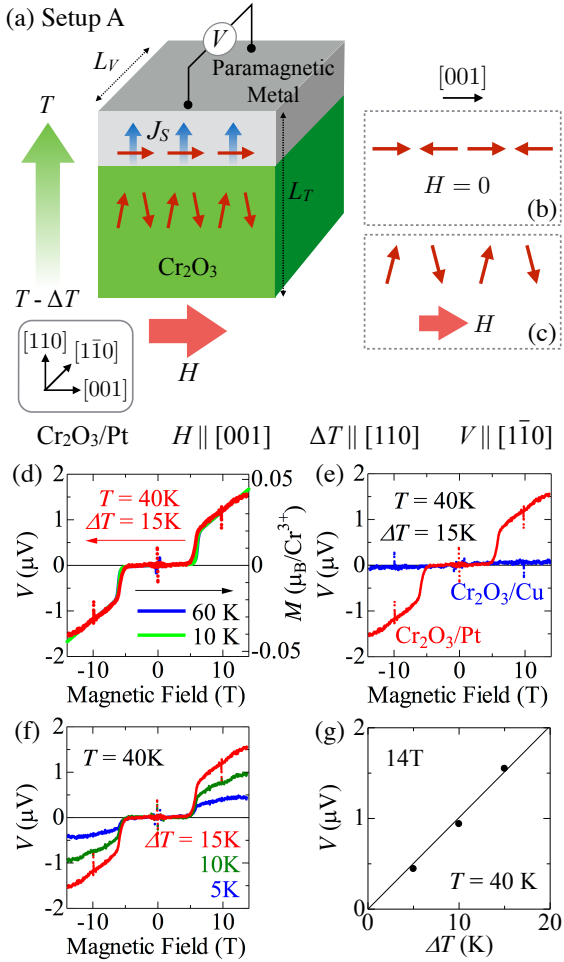}
\caption{(color online). (a) Experimental setup for the measurement of longitudinal spin Seebeck effect, with magnetic field ($H$) applied along the [001] axis of Cr$_2$O$_3$. Arrows in Cr$_2$O$_3$ represent the local magnetic moments, and bold blue (thin red) arrows in paramagnetic metal correspond to the propagation direction (carried spin angular momentum) of associated spin current $J_s$. $L_T$ (thickness of bulk Cr$_2$O$_3$ along the temperature gradient direction) and $L_V$ (distance between the electrodes (black circles) on the metal layer) are 0.5 mm and 4 mm, respectively. Unless specified, Pt is employed as the paramagnetic metal. (b) and (c) represent the magnetic structures of Cr$_2$O$_3$ for the ground state (i.e. $H=0$) and the $H$-induced spin flopped state, respectively. (d) $H$-dependence of induced electric voltage $V$ for Pt, and magnetization $M$ for Cr$_2$O$_3$. The similar voltage profiles are also measured with (e) different paramagnetic metal (Cu) and (f) different magnitudes of temperature gradient $\Delta T$. (g) $\Delta T$ dependence of Pt voltage at 14 T.}
\label{SSE}
\end{center}
\end{figure}

Bulk single crystals of Cr$_2$O$_3$ were grown by the laser floating zone method\cite{LaserFZ}. They are cut into rectangular shape, and polished with diamond slurry and colloidal silica. On the polished surface of Cr$_2$O$_3$, thin film of Pt (10 nm) or Cu (20 nm) is deposited as the PM layer by radio-frequency sputtering method. To provide the appropriate temperature gradient $\nabla T$, the sample is sandwiched with a pair of Cu blocks (covered by thin Al$_2$O$_3$ film to guarantee the electrical insulation but with good thermal contact) under the high vacuum condition less than 10$^{-4}$ Torr. One Cu block serves as the thermal bath with temperature $T - \Delta T$, and another Cu block is equipped with a resistive heater to keep its temperature $T$. Their temperatures are monitored by cernox thermometers and Lakeshore 335 temperature controller. Here, the temperature gradient is given by $\nabla T = \Delta T / L_T$ with $L_T$ being the sample thickness along the temperature gradient direction. To evaluate the magnitude of thermally-induced $J_s$ through Eq. (\ref{ISHE}), $H$-dependence of raw electric voltage $V_{\textrm{raw}}$ are measured in the PM layer with and without $\nabla T$ by nanovoltmeter. After the subtraction of background (i.e. the one with $\Delta T = 0$), the $H$-odd component of induced voltage $V$ is extracted by $V(H, \Delta T) = [(V_{\textrm{raw}}(H, \Delta T) - V_{\textrm{raw}}(H, 0))-(V_{\textrm{raw}}(-H, \Delta T) - V_{\textrm{raw}}(-H, 0))]/2$. Magnetization $M$ and thermal conductivity $\kappa$ for Cr$_2$O$_3$ are measured with Physical Properties Measurement System (PPMS, Quantum Design Inc).

The target compound Cr$_2$O$_3$ has corundum crystal structure with trigonal space group $R\bar{3}c$. The magnetism is dominated by the Cr$^{3+}$ ion with $S=3/2$, and the antiferromagnetic order with local magnetic moments pointing along the [001] axis is stabilized below the N\'{e}el temperature $T_{\rm{N}} \sim 308$ K (Fig. \ref{SSE} (b)). Since antiferromagnetically aligned spins prefer to lie normal to $H$, the application of $H \parallel [001]$ larger than critical field value $H_c$ induces spin-flop transition and reorients the magnetic moment direction as shown in Fig. \ref{SSE} (c)\cite{Cr2O3_Domain, Cr2O3_PhaseDiagram}. 

In the following, we mainly discuss the results for the Cr$_2$O$_3$/Pt sample under the experimental configuration shown in Fig. \ref{SSE} (a) (i.e. setup A) unless specified. Here, Pt is deposited on the (110) plane of Cr$_2$O$_3$ and $\nabla T$ is applied normal to it, which corresponds to the geometry of longitudinal spin Seebeck effect\cite{FirstLongSSE, YIG_SSE_NoProx}. Magnetic field is applied along the [001] direction of Cr$_2$O$_3$. To detect the electric voltage of spin current origin following Eq. (\ref{ISHE}), $V$ component normal to $H$ is measured within the Pt layer. Figure \ref{SSE} (d) indicates the magnetic field dependence of $M$ for Cr$_2$O$_3$, as well as $V$ in the Pt layer at $T=40$K and $\Delta T=15$ K. The application of $H \parallel [001]$ larger than $H_c \sim 6$ T causes a spin-flop transition and magnetization step in the $M$-$H$ profile, which remains almost $T$-independent below 60 K. Correspondingly, a clear step-like enhancement of $V$ is observed at $H_c$. The magnitude of $V$ in Pt is found to be proportional to $M$ in Cr$_2$O$_3$, suggesting that the observed voltage originates from thermally-induced spin current mediated by antiferromagnetic spin wave carrying nonzero spin angular momentum $\sigma \propto M$.

To further establish the validity of Eq. (\ref{ISHE}) in this system, the same voltage measurement is performed for the Cr$_2$O$_3$/Cu sample (Fig. \ref{SSE} (e)). The obtained $V$ in the Cu layer is negligibly small, consistent with the much smaller spin Hall angle for Cu ($\theta_{\rm SH} \sim 0.003$) than that for Pt ($\theta_{\rm SH} \sim 0.1$)\cite{SpinHallAngle}. The measurements are also performed under different magnitudes of $\Delta T$ for the Cr$_2$O$_3$/Pt sample while keeping $T=40$ K (Fig. \ref{SSE} (f)). Figure \ref{SSE} (g) summarizes the $\Delta T$-dependence of $V$-value at $H=14$ T, and $V$ turns out to be proportional to $\Delta T$. Previously, the relationship $J_s \propto \nabla T$ has been proposed for several ferromagnetic materials such as YIG\cite{FirstLongSSE}, and our present results suggest that it also holds for antiferromagnets.

\begin{figure}
\begin{center}
\includegraphics*[width=8cm]{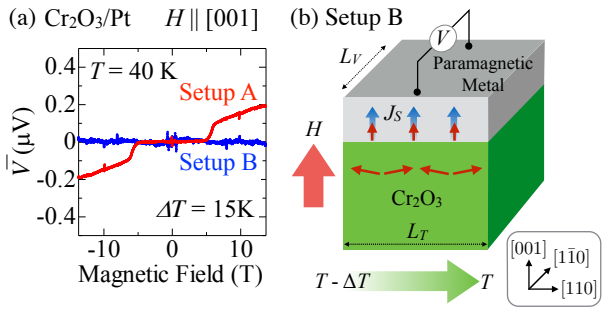}
\caption{(color online). (a) Magnetic field dependence of the normalized Pt voltage $\bar{V}$ ($=VL_T/L_V$) for $H$ applied along the [001] axis of Cr$_2$O$_3$, measured with the setup A (i.e. Fig. \ref{SSE} (a)) and the setup B as shown in (b). For the latter case, $L_T$ and $L_V$ are 2 mm and 4 mm, respectively, and only the anomalous Nernst effect can contribute to the induced voltage.}
\label{NoNernst}
\end{center}
\end{figure}

For the YIG/Pt system, the possible contribution of anomalous Nernst effect (ANE) into the $H$-odd voltage component has recently been discussed\cite{PtProximity, YIG_SSE_NoProx}. This scenario assumes the proximity ferromagnetism in the Pt layer (with local magnetization $M_{\rm Pt}$) at the interface with YIG, and the ANE contribution to the voltage is given by
\begin{equation}
\vec{V}_{\rm ANE} \propto L_V (\vec{M}_{\rm Pt} \times \vec{\nabla} T).
\end{equation}
In case of the setup A (Fig. \ref{SSE} (a)), the observed voltage comprises $V=V_{\rm ISHE} + V_{\rm ANE}$ and thus the subtraction of $V_{\rm ANE}$ is necessary to extract the pure contribution of $V_{\rm ISHE}$. For this purpose, we employed the different experimental setup as shown in Fig. \ref{NoNernst} (b) (i.e. setup B). Here, Pt is deposited on the (001) plane of Cr$_2$O$_3$ and $H$ is applied perpendicular to it. $\nabla T$ is along the in-plane [110] direction, and voltage component normal to both $\nabla T$ and $H$ is measured within the Pt layer. In this setup B, $V_{\rm ISHE}$ becomes zero due to $\vec{J}_s \parallel \vec{\sigma}$ and only $V_{\rm ANE}$ can contribute to the observed voltage\cite{YIG_SSE_NoProx}. In Fig. \ref{NoNernst} (a), $H$-dependence of $\bar{V}$ ($=V L_T / L_V$), i.e. voltage normalized with sample dimensions, is plotted for the Cr$_2$O$_3$/Pt sample with both setups A and B. While $H$ induces the spin-flop transition at 6 T in Cr$_2$O$_3$ for both configurations, the discernible voltage signal in Pt is observed only for the setup A. This proves that the contribution of $V_{\rm ISHE}$ is dominant and $V_{\rm ANE}$ is negligibly small in the present sample. In case of YIG/Pt with the same $\Delta T$ value, the inverse spin Hall voltage of $\bar{V} \sim 3 \mu$V has been reported in the saturated ferrimagnetic state at room temperature\cite{YIG_SSE_NoProx}. The presently observed $\bar{V} \sim 0.2 \mu$V for Cr$_2$O$_3$/Pt at 14 T is comparable with the above value, considering the relatively small magnitude of induced $M$ (less than 2 \% of saturated magnetization) and the $\bar{V} \propto M$ relationship observed for the thermal excitation process (Fig. \ref{SSE} (d)).

\begin{figure}
\begin{center}
\includegraphics*[width=8cm]{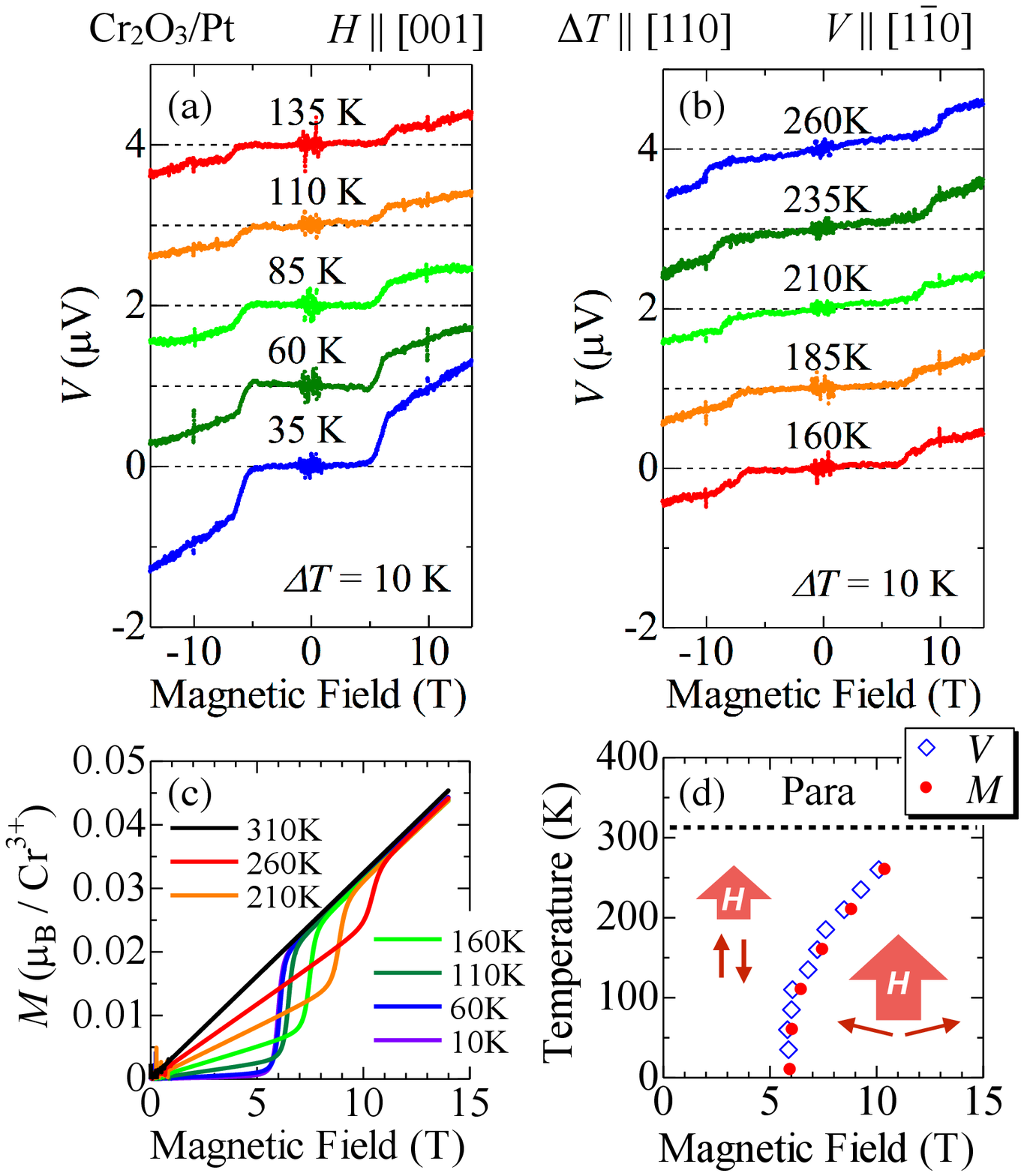}
\caption{(color online). (a) and (b) Magnetic field dependence of Pt voltage measured at various temperatures $T$ for the setup A (i.e. Fig. \ref{SSE} (a)), with the constant temperature gradient $\Delta T = 10$ K. (c) The corresponding magnetic field dependence of magnetization for Cr$_2$O$_3$ with $H \parallel [001]$. (d) $H$-$T$ phase diagram for Cr$_2$O$_3$ with $H \parallel [001]$, deduced from anomalies in magnetization profile (red closed circles). Above $T_{\rm{N}} \sim 308$ K, Cr$_2$O$_3$ becomes paramagnetic. The step-like anomalies for Pt voltage observed in (a) and (b) are also plotted as blue open diamonds.}
\label{Tdep}
\end{center}
\end{figure}

\begin{figure}
\begin{center}
\includegraphics*[width=8cm]{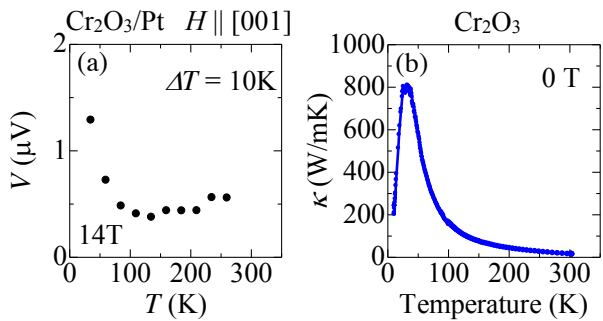}
\caption{(color online). (a) Temperature dependence of inverse spin Hall voltage for the Cr$_2$O$_3$/Pt system at 14 T, obtained from the data in Figs. \ref{Tdep} (a) and (b). (b) Temperature dependence of thermal conductivity $\kappa$ along the [110] axis for Cr$_2$O$_3$, measured under zero magnetic field.} 
\label{PhononDrag}
\end{center}
\end{figure}

Next, we investigated the temperature dependence of voltage profiles for the Cr$_2$O$_3$/Pt sample with the original setup A. Figures \ref{Tdep} (a) and (b) indicate the $H$-dependence of $V$ in the Pt layer, measured at various temperatures keeping $\Delta T = 10$K. The corresponding $H$-dependence of $M$ for Cr$_2$O$_3$ obtained at various $T$ is also plotted in Fig. \ref{Tdep} (c). Both $V$ for Pt and $M$ for Cr$_2$O$_3$ show a clear step-like anomaly corresponding to the spin-flop transition at $H_c$. Based on these measurements, $H$-$T$ magnetic phase diagram for Cr$_2$O$_3$ under $H \parallel [001]$ is summarized in Fig. \ref{Tdep} (d)\cite{Cr2O3_PhaseDiagram}. While $H_c$ becomes larger for higher temperature, the anomalies in $V$ and $M$ always coincide with each other. This confirms that the observed $V$ in Pt clearly reflects the magnetic nature of underlying Cr$_2$O$_3$. In Fig. \ref{PhononDrag} (a), the magnitude of $V$ obtained at 14 T is plotted as a function of temperature. The $V$-value shows clear enhancement for lower $T$, while the corresponding $M$-value at 14T remains almost unchanged for whole temperature range as shown in Fig. \ref{Tdep} (c). Similar increase of $V$ for lower $T$ has recently been reported for YIG/Pt, where the relevance of phonon-drag effect has been proposed\cite{YIG_PhononDrag, UchidaReview, GaMnAs_PhononDrag}. This scenario assumes that thermally-induced propagating phonons drag magnons through magnon-phonon interaction, and the prolonged phonon-lifetime $\tau_{\rm{ph}}$ in lower-$T$ region enhances the magnitude of $J_s$. For insulating materials, the $T$-dependence of $\tau_{\rm{ph}}$ can be estimated from thermal conductivity $\kappa$ using the relationship $\kappa = v_{\rm{ph}}^2 C_{\rm{ph}} \tau_{\rm{ph}}$/3\cite{Mermin_Text, Cr2O3_Comment}. Here, $v_{\rm{ph}}$ represents $T$-independent phonon group velocity, and $C_{\rm{ph}}$ does phonon heat capacity whose magnitude monotonically decreases for lower $T$. Figure \ref{Tdep} (d) indicates temperature dependence of $\kappa$ along the [110] axis measured for Cr$_2$O$_3$. $\kappa$ shows clear enhancement for lower-$T$ and has maximum around 30 K, which reflects the elongation of $\tau_{\rm{ph}}$ for lower $T$ due to the suppression of Umklapp phonon scattering. The present results strongly suggest that the phonon-drag process and the resultant enhancement of $J_s$ also take place in antiferromagnets. Note that the $V$-value in Pt shows slight upturn above 120 K, which may originate from the increase of magnon population.

In general, simple uniaxial antiferromagnets are characterized by two degenerated magnon branches with dispersion relationship $\nu^+ (k)$ and $\nu^- (k)$, which are expected to carry opposite sign of spin angular momentum\cite{AF_SpinPump, AF_SSE}. This degeneracy is lifted under $H$ applied along the magnetic easy axis, where $\nu^+ (k)$ ($\nu^- (k)$) linearly increases (decreases) as a function of $H$. When $\nu^- (k)$ reaches zero at $H = H_c$, the spin-flopped state characterized by two different magnon branches carrying the same sign of spin angular momentum is stabilized\cite{AFMR_Text}. Since the thermal process excites magnon modes of any wavenumber $k$ and frequency $\nu$ following the Bose distribution function, $\nabla T$-induced generation of nonzero spin current is always justified for antiferromagnets with finite $H$. Note that the typical eigen frequency of antiferromagnetic spin wave ranges from the order of a hundred gigahertz to terahertz, which is much higher than that of ferromagnetic one in the order of gigahertz\cite{AFMR_Text}. In case of Cr$_2$O$_3$, antiferromagnetic resonance frequency 170 GHz $(= \nu^{\pm}(0))$ has previously been reported for the ground state\cite{Cr2O3_AFMR, Cr2O3_AFMR2}. Considering that 1 K of thermal fluctuation corresponds to 20.8 GHz of photon frequency, such antiferromagnetic spin waves can be easily excited through the thermal process for the presently employed temperature range. Recent theories predict that each of two degenerated antiferromagnetic magnon branches for $H \rightarrow 0$ is also active for circularly polarized microwave but with opposite handedness\cite{AF_SpinPump, NiO_spinpump}, and the efficient optical generation of spin current through selective mode excitation under zero magnetic field would be an interesting challenge for antiferromagnets.

In summary, we have experimentally observed longitudinal spin Seebeck effect for an uniaxial antiferromagnetic insulator Cr$_2$O$_3$. The application of temperature gradient normal to the Cr$_2$O$_3$/Pt interface causes the inverse spin Hall voltage of spin current origin in the Pt layer, whose magnitude turns out to be proportional to magnetization $M$ in Cr$_2$O$_3$. The present finding demonstrates that the high-frequency antiferromagnetic spin wave above 100 GHz can be an efficient carrier of spin current, which highlights antiferromagnetic insulators as the promising source of unique spintronic functions.

The authors thank T. Arima, K. S. Takahashi, Y. Iwasa, Y. Kasahara, H. Matsui, Y. Tokunaga, A. Kikkawa, Y. Ohnuma, Y. Shiomi, K. Uchida and E. Saitoh for enlightening discussions and experimental helps. This work was partly supported by the Mitsubishi Foundation, Grants-In-Aid for Scientific Research (Grant No. 26610109, 15H05458) from the MEXT of Japan, and FIRST Program by the Japan Society for the Promotion of Science (JSPS) .


\begin{thebibliography}{100}
\bibitem{Cr2O3_Comment} For Cr$_2$O$_3$, the previous thermal conductivity measurement has concluded that its heat transport is dominated by the contribution of phonons\cite{Cr2O3_Thermal}.
\end{thebibliography}
\end{document}